# Phase Field Study of Exchange Coupling of Hard/Soft Ferrite on Magnetic Permeability

Xinyu Xu[1, a)], Wenqin Yue[2, a)], Yueli Yu[3], Yongke Yan[4*], and Liwei D. Geng[2*]

[1] *Department of Materials Science and Engineering, Stanford University, Stanford, CA 94305, USA*
[2] *College of Materials Science and Engineering, Sichuan University, Chengdu, Sichuan 610065, P. R. China*
[3] *Department of Materials Science and Engineering, Northwestern University, Evanston, IL, 60208*
[4] *Electronic Materials Research Laboratory, Key Laboratory of the Ministry of Education and International Center for Dielectric Research, School of Electronic Science and Engineering, Xi'an Jiaotong University, Xi'an 710049, P.R. China*

(*Electronic mail: liwei.geng@scupi.cn (L.G.); yanyongke@xjtu.edu.cn (Y.Y.))

**Abstract**

Effective modulation of magnetic permeability plays a vital role in the development of high-performance inductors. Here, phase-field simulations of hard/soft ferrite composites (BaM/NiZn) clarify how exchange coupling and microstructure impact magnetic permeability. We show that particle size, volume fraction, and orientation of the hard phase can effectively control the transition from collinear to non-collinear coupling, with a critical exchange size $r_{cr} \approx 12$ nm. Increasing the hard-phase fraction deepens the anisotropy energy well and monotonically suppresses permeability. In contrast, rotating the BaM easy axis to 90° relative to the applied field produces a strong enhancement: at a 10 nm radius and $\eta = 0.1$ volume fraction, the effective permeability could be more than 30 times larger than in the parallel configuration, and then saturates for larger particles. This study establishes a microstructure–permeability–based physical framework for designing hard/soft magnetic composite systems.

**Introduction**

Magnetic materials play a vital role in high-frequency inductors and related magnetic devices, where the effective magnetic permeability is a key parameter that determines device size, loss, and frequency response[1,2]. Recent advances in material design, composite architectures, and inductor topology have improved device performance to some extent[3-8]. Nevertheless, these studies primarily focus on device-level metrics and rely heavily on empirical or semi-empirical choices of soft magnetic materials and structures[9,10]. A quantitative and predictive understanding of how the microstructure of the soft magnetic phase determines the effective permeability remains largely underdeveloped, limiting microstructure-driven and permeability-targeted design strategies for specific applications[11,12].

NiZn ferrite is widely used as a soft magnetic phase in high-frequency inductors due to its low energy loss[13-15]. However, in practical fabrication, particularly when co-firing is employed to enhance mechanical reliability and integration density, differences in thermal expansion and sintering shrinkage introduce structural mismatch and residual stress. These effects cause the effective permeability of the soft magnetic layer to deviate from its intended value and make it difficult to control with precision[5]. To retain the low-loss characteristics of NiZn ferrite while achieving more controllable and predictable magnetic permeability, an attractive approach is to incorporate a hard magnetic phase into the soft magnetic matrix and utilize exchange coupling to reshape the local magnetic field distribution and domain configuration[9-12,16,17]. BaM hexaferrite, with its large uniaxial anisotropy field and favorable high-frequency performance, is a widely studied hard magnetic candidate for such composite systems[14,15,18]. In BaM/NiZn hard/soft ferrite composites, understanding how the particle size, volume fraction, and crystallographic orientation of the hard phase regulate both the magnitude and trends of the permeability is essential for establishing a microstructure-informed physical description[9,10,17].

Motivated by these considerations, this work employs domain-level phase field modeling and computer simulation to investigate the influence of exchange coupling on the magnetic permeability of BaM/NiZn hard/soft ferrite compos- ites. We examine how the size, volume fraction, and crystal- lographic rotation of hard-magnetic particles affect the quasi- static permeability response under an applied field and the evolution of magnetic domain structures. Collinear and non- collinear coupling regimes are identified and analyzed, to- gether with their corresponding critical size conditions. In ad- dition, by examining the evolution of magnetic anisotropy en- ergy and the easy-axis orientation with increasing hard phase volume fraction, we elucidate how the hard phase modulates the magnetization and the rotational freedom of mag- netization in the soft NiZn ferrite. This study aims to estab- lish a

permeability–microstructure-based physical framework for the design of BaM/NiZn composite systems and to provide magnetic-design guidelines for subsequent structural and material optimization at the device level.

**Model**

In our theoretical framework, the total free energy of the magnetic system is formulated as a function of the magnetization $\mathbf{M}(\mathbf{r})$. This total free energy encompasses four primary contributions: the magnetocrystalline anisotropy (MCA) energy, the exchange energy, the magnetostatic energy and the Zeeman energy. The total free energy under external magnetic field $\mathbf{H}^{ext}$ is given by:

$$F = \int f_{ani}(\mathbf{m})d^3r + A\int |\nabla \mathbf{m}|^2 d^3r \\ + \frac{\mu_0}{2}\int \frac{d^3k}{(2\pi)^3} \left|\mathbf{n} \cdot \widetilde{\mathbf{M}}(\mathbf{k})\right|^2 - \mu_0 \int \mathbf{H}^{ext} \cdot \mathbf{M}(\mathbf{r})d^3r \quad (1)$$

where $\mathbf{m}$ is the normalized magnetization vector, $f_{ani}(\mathbf{m})$ is the magnetocrystalline anisotropy energy density, $A$ is the exchange stiffness constant and $\mu_0$ is the vacuum permeability.

In this model, two types of magnetocrystalline anisotropy energy density are employed according to the crystalline structure of the materials. For the soft magnetic NiZn ferrite with cubic symmetry, the anisotropy energy density is given by[19]:

$$f_{ani}^{(soft)}[m] = K_c(m_1^2 m_2^2 + m_1^2 m_3^2 + m_2^2 m_3^2) \quad (2)$$

where $K_c$ is the cubit anisotropy constant.

For the hard magnetic phase, a uniaxial anisotropy model is used, with its energy density expressed as:

$$f_{ani}^{(hard)}[m] = K_u sinv \quad (3)$$

where $K_u$ is the magnetocrystalline anisotropy constant and $v$ is the angle between the magnetization direction and the anisotropy axis[19].

For the composite material consisting of a soft and hard magnetic, the effective anisotropy energy density is calculated as a weighted average based on the volume fractions. Defining $\eta$ as the volume fraction of the hard magnetic phase, the total energy density is given by:

$$f_{ani}[m] = (1-\eta)f_{ani}^{(soft)}[m] + \eta f_{ani}^{(hard)}[m] \quad (4)$$

To investigate the magnetization evolution in composites composed of single-crystal NiZn ferrite and hard-magnetic BaM particles, the micromagnetic phase-field model is employed[20]. The magnetic domain structure is represented by the vector field $\mathbf{M}(\mathbf{r})$, describing the spatial distribution of magnetization. Since $|\mathbf{M}| = M_s$ is constant, we define the normalized magnetization to characterize domain structures and their temporal evolution.

$$m(\mathbf{r}) = \frac{\mathbf{M}(\mathbf{r})}{M_s} \quad (5)$$

where $M_s$ is the saturated magnetization. The dynamic evolution of $\mathbf{m}$ is governed by the Landau–Lifshitz–Gilbert (LLG) equation[20,21]:

$$\dot{m} = -\gamma \mathbf{H}_{eff} \times m + \alpha m \times \dot{m} \quad (6)$$

where $\gamma$ is the gyromagnetic ratio, $\alpha$ is the damping parameter, and $\mathbf{H}_{eff}$ is the effective magnetic field which is derived from the total free energy equation[20]:

$$\mathbf{H}_{eff} = -\frac{1}{\mu_0} \frac{\delta F}{\delta \mathbf{M}} \quad (7)$$

The material parameters used in this study are $K_c = -1.5 \times 10^4$ J/m³ and $K_u = -3.3 \times 10^5$ J/m³. The saturation magnetizations are $M_s^{NiZn} = 3.5 \times 10^5$ A/m and $M_s^{BaM} = 3.8 \times 10^5$ A/m. For simplicity, we normalize the saturation magnetization by the smaller value, $M_s = 3.5 \times 10^5$ A/m. The exchange stiffness is set to $A = 1.0 \times 10^{-11}$ J/m. We employ a computational cell of size $256 \times 256 \times 1$ with periodic boundary conditions to model the magnetization evolution. The computational grid spacing is $\Delta x = 2.0$ nm.

**Result**

To elucidate the incorporation mechanism of hard magnetic particles into the soft magnetic phase, three key variables are considered in this study: the particle size, volume fraction, and crystallographic orientation of the hard magnetic particles. Hard magnetic particles are generally assumed to exhibit a spherical morphology; therefore, the particle size is conveniently characterized by the sphere radius. In the simulations, the soft magnetic phase is modeled as a single crystal. Figure 1(a) illustrates one representative phase morphology of the hard/soft magnetic composite. The easy axis of NiZn ferrite is aligned along the [111] direction, whereas that of BaM is aligned along the [001] direction, as indicated in Figure 1(b). Figure 1(c) shows the corresponding stabilized magnetic domain structure at equilibrium.

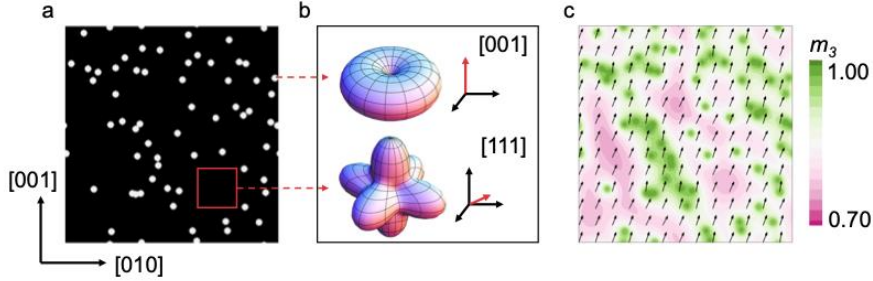

FIG. 1. Hard/soft magnetic composite with $\eta = 0.05$ BaM ($r$ = 8 nm). (a) Phase morphology of the hard/soft magnetic composite, where hard particles (white) are embedded within soft ferritematrix (black); (b) Anisotropy energy density for hard and soft ferrite with their easy axes. (c) Magnetic domain structure, where arrows indicate magnetization components along the [001] and [010] directions, and color contours represent the magnetization component along the [001] direction. The simulated region size is $521\ nm \times 521\ nm$.

To examine the effect of hard magnetic particle size, particles with radii ranging from 6 to 20 nm were analyzed, while both the hard magnetic phase and the NiZn ferrite phase were aligned along the same crystallographic orientation. After the magnetization reached equilibrium, a small external magnetic field was applied along the [001] direction.

Figure 2(a) shows the variation in permeability as a function of particle size. Under strong interaction conditions, the magnetizations of the hard and soft magnetic phases remain essentially collinear. The influence range of the strong interaction, characterized by the critical dimension $r_{cr}$, can be evaluated by,

$$r_{cr} \approx \pi \sqrt{\frac{A}{2K}} \quad (8)$$

and the exchange interaction operates over $r_{cr} \approx 12$ nm. In this collinear regime, the permeability increases steadily from 1.4 to 2.4 as the particle radius increases. This trend arises because smaller particles possess a larger surface-to-volume ratio, which enhances their influence on the magnetization distribution in the soft magnetic phase. Consequently, the magnetization of the soft phase tends to align more strongly with the [001] easy axis of the hard magnetic particles. Since the external field is applied along the same direction (Figure 2(c)), the NiZn ferrite phase experiences limited room for magnetization rotation, resulting in reduced permeability.

Figure 2(b) further confirms that smaller particle sizes yield a larger magnetization component along the [001] direction. Once the particle size exceeds the regime dominated by interfacial effects, the permeability curve saturates, indicating that when the particle size exceeds $r_{cr}$, the exchange coupling is no longer sufficient to maintain collinearity, and the hard and soft phases can be regarded as entering a non-collinear regime.

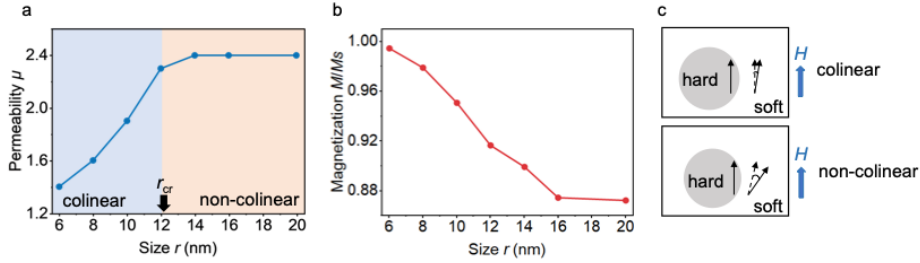

FIG. 2. Size effect of hard magnetic particles on (a) permeability, and (b) magnetization along [001] direction with volume fraction of $\eta = 0.1$. (c) Schematic illustration of magnetization couplings in hard/soft magnetic composite.

Figure 3 illustrates the magnetic domain structures and the impact of doping hard magnetic particles into the soft magnetic phase. In Figure 3(a), where the hard magnetic particles have a radius of 6 nm, the composite exhibits a distinctly collinear configuration. In this case, there is very little room for the magnetization to respond to an external magnetic field, leading to a lower permeability, consistent with the trend in Figure 2. As the size of the hard magnetic particles increases, their constraint on the orientation of the soft phase magnetization is gradually reduced, which allows the permeability to increase. In particular, Figure 3(d) shows a clear change in magnetization at the interface between the soft and hard magnetic phases. The magnetization in the soft magnetic phase gradually rotates from a near-[111] orientation toward the [001] direction, clearly indicating a strong coupling between the two phases. Furthermore, Figures 3(e) and 3(h) indicate that the influence of the hard magnetic particles is essentially limited to a local zone extending only a distance $r_{cr}$ into the soft phase, while beyond this range their effect on the soft magnetic matrix is much weaker. This spatially localized response explains why, in the weak-coupling regime, the overall impact of the hard magnetic particles on the soft magnetic phase is limited, corresponding to the leveling-off behavior observed in Figure 2.

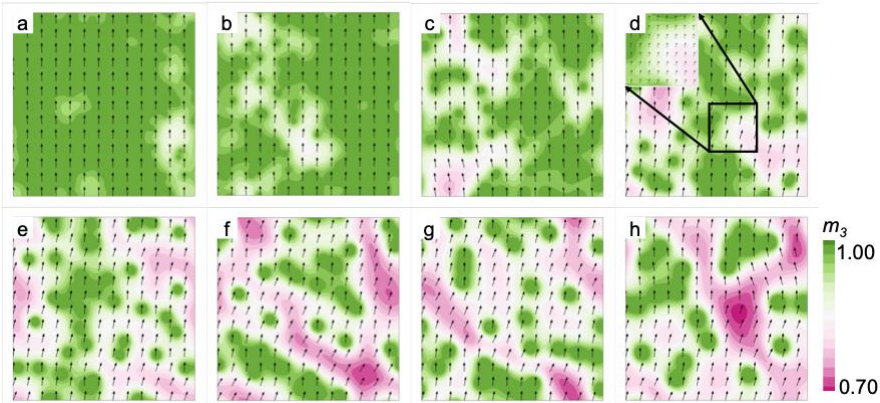

FIG. 3. Simulated magnetization distributions of hard/soft ferrite composites with particle sizes of (a) 6, (b) 8, (c) 10, (d) 12, (e) 14, (f) 16, (g) 18 and (h) 20nm with volume fraction of $\eta = 0.1$.

Subsequently, we examine the relationship between the volume fraction of hard magnetic particles and the permeability. Figure 4 presents the variations in permeability and the magnetization component along the [001] direction as the hard phase fraction increases. As shown in Figure 4(a), the permeability decreases monotonically with increasing volume fraction for all particle radii. For instance, at a volume fraction of 0.03 the permeability lies in the range 3.0–4.4 depending on particle size, but it drops to about 1.4–2.4 when the fraction increases to 0.10. Figure 4(b) shows the corresponding magnetization magnitude along the [001] direction. These curves display an inverse relationship with permeability: as the available space for free alignment in the soft phase diminishes, a larger portion of the magnetization is forced to align along [001], increasing the net component in this direction. These trends reflect the increasing influence of the hard magnetic phase on the soft magnetic phase as its volume fraction grows. As more regions of the soft phase become pinned or constrained by the hard particles, the local magnetization more readily reorients toward [001], reducing the freedom for magnetization rotation and thereby lowering the permeability. Figure 5 illustrates, under strong interaction conditions, how increasing the volume fraction of hard magnetic particles modifies the local easy-axis distribution in the soft phase. As the hard-phase fraction increases, the preferred direction in the NiZn ferrite gradually shifts from [111] toward [001], and at a hard-phase volume fraction of 0.05 the local easy axis becomes fully aligned with [001].

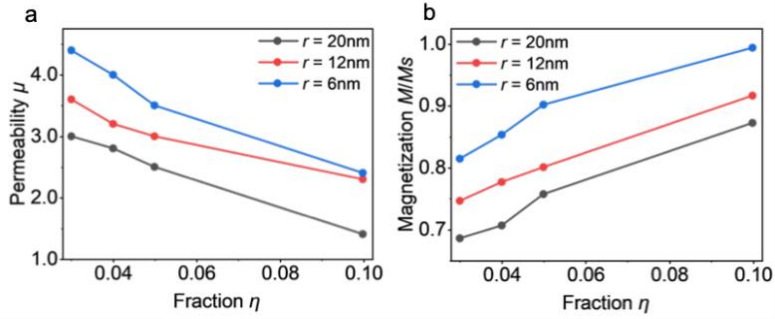

FIG. 4. Simulated (a) permeability and (b) magnetization along [001] direction as a function of volume fraction $\eta$ with various particle sizes.

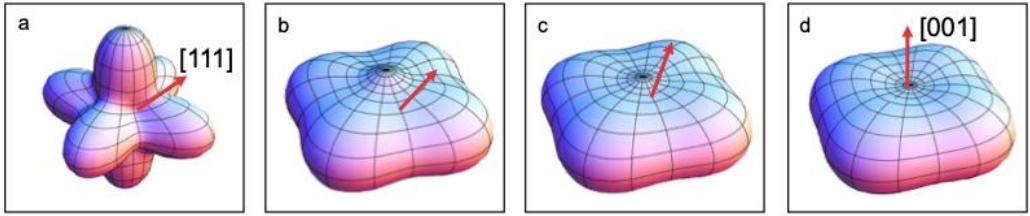

FIG. 5. Magnetocrystalline anisotropy energy contours of hard/soft ferrite composites with increasing hard phase volume fraction $\eta$: (a) 0, (b) 0.03, (c) 0.04, and (d) 0.05. Red vector represents the easy axis.

Figure 6 illustrates the influence of increasing the volume fraction of the hard magnetic phase on the magnetic domain structure. As the hard-phase volume fraction increases, the coupling between the hard and soft phases is strengthened. This is evidenced by more pronounced local domain walls and an expanded interfacial region between the two phases. Because a larger proportion of the soft-phase magnetization is reoriented toward the [001] easy axis of the hard phase, the regions corresponding to the soft magnetic phase (in pink) become more fragmented, while the regions where the hard phase dominates the local magnetization (in green) expand correspondingly. In Figures 6(c), 6(f), and 6(i), where the hard-phase volume fraction reaches 0.05, this coupling effect is most pronounced. The soft magnetic regions appear more discontinuous, and the magnetization arrows tend to align more uniformly along the [001] direction. As an increasing fraction of the soft-phase magnetization becomes constrained along [001], the number of moments that can freely rotate is reduced. This leads to a more restricted magnetization process under an applied field and, consequently, a decrease in the effective permeability.

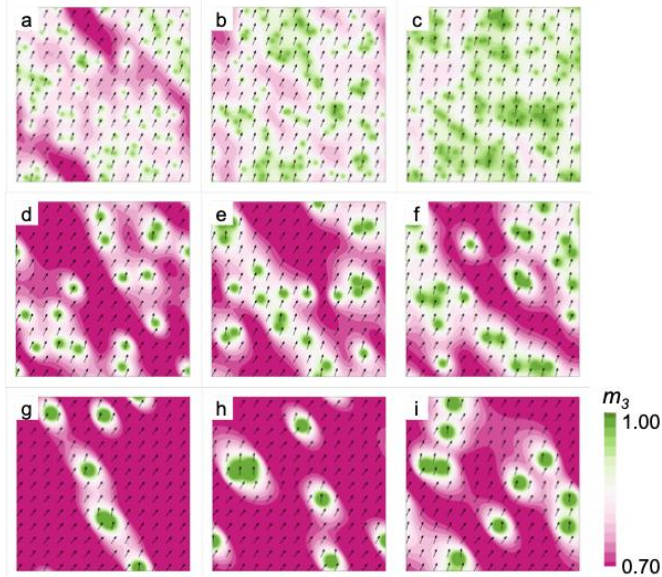

FIG. 6. Simulated magnetization distributions of hard/soft ferrite composites for various hard-phase volume fractions $\eta$ and sizes $r$: (a) $\eta = 0.03$, (b) 0.04, (c) 0.05 with $r = 6$ nm; (d) $\eta = 0.03$, (e) 0.04, (f) 0.05 with $r = 12$ nm; (g) $\eta = 0.03$, (h) 0.04, (i) 0.05 with $r = 20$ nm.

The orientation of the hard magnetic particles is particularly critical, as it directly determines the direction of the easy axis. Because the magnetic anisotropy energy governs both domain-wall motion and magnetization rotation, it strongly influences the magnetic response and permeability of hard/soft magnetic composites. The orientation of the hard magnetic particles is varied by rotating their uniaxial easy axis, which contains no magnetization components along the other two crystallographic directions. This property allows clear observation and control of the magnetization orientation. Since the easy axis of the hard phase lies along [001], rotation about this axis would not produce any significant change. Therefore, the [100] axis—perpendicular to the magnetic domain structure plane—was selected as the rotation axis to enable direct comparison and visualization of domain evolution. Rotation angles of 15°, 30°, 45°, 60°, 75°, and 90° were applied while keeping the particle radius fixed at 20 nm and the volume fraction at 0.1. Among these cases, the 90° rotation represents a critical configuration for examining the interplay between particle orientation and permeability.

Figure 7 shows that introducing hard magnetic particles with different orientations significantly affects the magnetic permeability. When the orientation angle is 0°, meaning the external field is parallel to the easy axis of the hard phase, the permeability remains relatively low. Even a slight deviation from this parallel alignment produces a noticeable increase, underscoring the sensitivity of the composite to particle orientation. As the orientation angle increases, the permeability continues to rise steadily until the configuration approaches the perpendicular case at 90°. At this orientation, a sharp jump in permeability occurs, resulting in a value much higher than that of the parallel case. This abrupt increase indicates that the soft magnetic phase gains substantially greater rotational freedom once it is no longer constrained by the hard phase's parallel alignment, thereby enhancing the overall magnetic permeability of the composite.

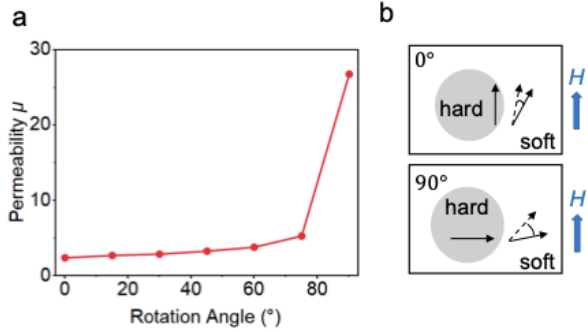

FIG. 7. Simulated magnetic permeability with increasing rotation angle $\theta$ of hard ferrite with $r = 20$ nm and $\eta = 0.1$. (b) Schematic illustration of magnetization couplings in hard/soft magnetic composite.

Figure 8 presents the magnetic domain structures in detail and clarifies the mechanism by which particle orientation affects the permeability. In this case, the easy axis of the hard magnetic particles deviates from the original [001] direction by approximately 15°. As a result, the NiZn ferrite phase exhibits a noticeable tilt away from [001], as shown in Figure 8(a), developing a stronger magnetization component toward the [010] direction. This behavior demonstrates that strong coupling with the hard magnetic phase effectively reorients the soft magnetic domains along the direction favored by the hard phase. As the rotation angle of the hard magnetic particles increases, the magnetization within the NiZn ferrite progressively shifts from predominantly [001] toward [010], as illustrated in Figures 8(b)–8(e). At a rotation angle of 90°, the soft magnetic domain orientations align predominantly along [010], as shown in Figure 8(f), reflecting substantial alignment with the easy axis of the hard magnetic phase in this configuration. This progressive reorientation of domains provides greater freedom for subsequent rotation toward [001] when an external magnetic field is applied along the [001] axis, thereby enhancing the magnetic permeability.

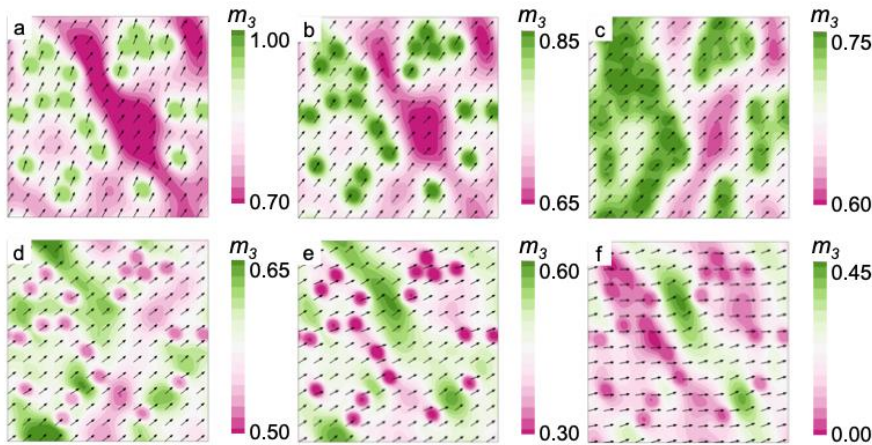

FIG. 8. Simulated magnetization distributions of hard/soft ferrite composites with rotation angles of a) $\theta=15°$, b) 30°, c) 45°, d) 60°, e) 75°, and f) 90° under $r = 20$ nm and $\eta = 10\%$.

Next, we examine how the particle size of the hard magnetic phase influences the magnetic permeability at an orientation of 90°. As shown in Figure 9, the permeability behavior in this configuration exhibits a markedly non-monotonic trend. At 90°, the overall permeability is substantially enhanced, reaching a peak value that is approximately 33 times larger than that in the parallel (0°) configuration. This dramatic increase arises from the dominant role of the hard magnetic phase when its easy axis is nearly perpendicular to the [001] direction. In this case, the [001] component of the hard-phase magnetization is nearly zero, so even a small external field applied along [001] can induce a large reorientation of the composite magnetization, resulting in an exceptionally high permeability. For hard magnetic particles with radii below 10 nm, the permeability increases with particle size and reaches a maximum of about 63 at a radius of 10 nm. Beyond this radius, the permeability decreases gradually with further increases in particle size and eventually stabilizes within a range of approximately 20–30.

This behavior originates from the strong interaction between the hard magnetic particles and the NiZn ferrite phase. For smaller hard particles, the interfacial coupling is stronger and their influence on the local internal field is more pronounced, which makes the magnetization of the NiZn ferrite more resistant to reorientation under an external field and thereby suppresses magnetization rotation. The permeability reaches a maximum at a particle radius of 10 nm, corresponding to an optimal balance: the hard phase is still sufficiently influential to induce a large rotation of the NiZn ferrite magnetization, yet not so dominant as to overconstrain its mobility. Above this critical dimension, the effective coupling from the hard magnetic particles weakens markedly, leading to a rapid decrease in permeability. As the particle size continues to increase beyond this threshold, the influence of the hard phase on the NiZn ferrite magnetization becomes marginal, and the permeability eventually saturates.

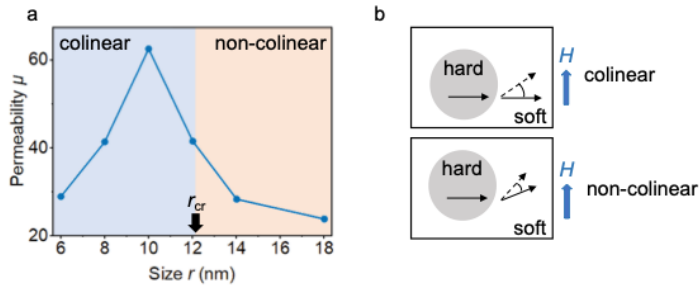

FIG. 9. Simulated magnetic permeability as a function of hard-phase particle size r with $\eta = 0.1$ and $\theta = 90°$. (b) Schematic illustration of magnetization couplings in hard/soft magnetic composite.

Figure 10 illustrates the evolution of the magnetic domain structures and local magnetic moment distributions in the composite system at a rotation angle of 90° as the BaM hard magnetic particle radius increases from 6 nm to 18 nm and 20 nm. In Figure 10(a), for a particle radius of 6 nm, most arrows are oriented close to the [010] direction, which becomes the easy axis after the hard phase is rotated by 90°. This configuration reflects strong coupling between the hard and soft phases: the soft magnetic phase is tightly constrained to follow the hard phase, and its ability to respond to an external field is strongly suppressed, leading to reduced permeability. In Figure 10(b), regions with lighter color contrast indicate a modest increase in the [001] magnetization component, and the arrows are slightly more dispersed, with some moments tilting toward [001]. This suggests that, as the particle size increases, the interfacial coupling remains strong, but the constraint imposed on soft magnetic regions farther away from the hard particles becomes weaker, thereby allowing a larger [001] component to develop. Figure 10(c) corresponds to an intermediate particle size, where the system attains a relatively balanced state. The influence of the hard magnetic particles on the soft phase is still sufficient to produce a clear [010] magnetization component, yet the ability of the hard phase to pin the soft phase is reduced. As a consequence, the soft magnetic domain structure retains considerable rotational freedom. This intermediate size yields an optimal compromise between specific surface area, interfacial coupling, and magnetic mobility, and is consistent with the peak permeability observed in Figure 9. For larger particles, as shown in Figures 10(d)–10(f), many arrows deviate more substantially from the [010] direction. This behavior indicates that the effective coupling between the soft and hard phases deteriorates as particle size continues to increase. The overall coupling efficiency of the composite is reduced, and the magnetization becomes less adaptable to the external magnetic field, as evidenced by the permeability trend in Figure 9(f). Taken together, these results reinforce the concept of an optimal particle size. Intermediate particle sizes provide a delicate balance between strong coupling and sufficient rotational freedom in the soft magnetic phase, resulting in a maximum permeability. In contrast, deviations toward smaller sizes overconstrain the local magnetization, while larger sizes diminish the overall coupling effectiveness, both leading to a reduction in magnetic permeability.

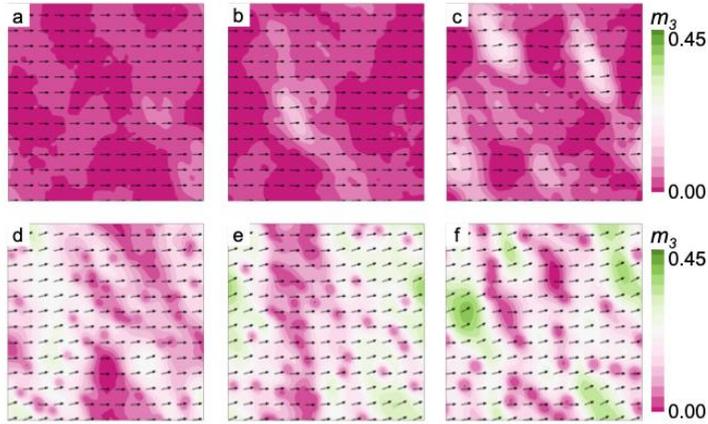

FIG. 10. Simulated magnetization distributions of hard/soft ferrite composites with particle sizes a) 6, b) 8, c) 10, d) 12, e) 14, and f) 18 nm under $\eta = 0.1$ and $\theta = 90°$.

**Conclusion**

In summary, this work clarifies how exchange coupling between hard BaM and soft NiZn ferrite governs the magnetic permeability of composite systems. Phase-field simulations show that the particle size, volume fraction, and orientation of the hard phase jointly tune the balance between interfacial coupling and the rotational freedom of the soft phase. A clear transition from a collinear to a non-collinear coupling regime is identified. For particle radii below the critical exchange dimension $r_{cr} \approx 12$ nm, the magnetizations of the hard and soft phases remain essentially collinear, and strong coupling constrains soft-phase rotation, limiting the permeability. Above $r_{cr}$, the coupling becomes more localized near interfaces, the soft phase away from the particles can reorient more independently, and the permeability response is accordingly modified. The results further demonstrate that increasing the hard-phase volume fraction enhances the effective magnetic anisotropy and suppresses permeability, whereas rotating the BaM easy axis away from [001] can dramatically enhance it. In particular, at a 90° orientation, the permeability exhibits a pronounced maximum at an intermediate particle size on the order of 10 nm, exceeding the parallel case by more than an order of magnitude. These findings highlight the existence of an optimal combination of particle size, volume fraction, and orientation that balances collinear and non-collinear exchange coupling, providing useful guidelines for the design of high-permeability hard/soft ferrite composites and voltage tunable inductors.

**Acknowledgement**

This work was supported by the National Natural Science Foundation of China (No. 52272120, 52032010), and the National Key Research and Development Program of China (No. 2023YFF0720700). The computer simulations were performed at the Hefei Advanced Computing Center.

**Data Availability Statement**

Data sharing is not applicable to this article as no new data were created or analyzed in this study.